\documentclass[twocolumn,letterpaper,aps,prl,superscriptaddress,showpacs,floatfix]{revtex4}

\usepackage{graphicx}	% Include figure filed
\usepackage{xspace}	% Include xspace

\newcommand{\pp}{\mbox{$p$+$p$}\xspace}
\newcommand{\pt}{\mbox{$p_T$}\xspace}

\newcommand{\raa}{\mbox{$R_{\rm AA}$}\xspace}

\newcommand{\sqsn}{\mbox{$\sqrt{s_{_{NN}}}$}\xspace}
\newcommand{\gevc}{\mbox{GeV/$c$}\xspace}
\newcommand{\piz}{\mbox{$\pi^0$}\xspace}
\newcommand{\mean}[1]{\left\langle #1 \right\rangle}

\begin{document}

\title{Measurement of Direct Photons in Au$+$Au Collisions at $\sqrt{s_{_{NN}}}$ = 
200 GeV}

\newcommand{\abilene}{Abilene Christian University, Abilene, Texas 79699, USA}
\newcommand{\banaras}{Department of Physics, Banaras Hindu University, Varanasi 221005, India}
\newcommand{\bnlphys}{Brookhaven National Laboratory, Upton, New York 11973-5000, USA}
\newcommand{\caucr}{University of California - Riverside, Riverside, California 92521, USA}
\newcommand{\cns}{Center for Nuclear Study, Graduate School of Science, University of Tokyo, 7-3-1 Hongo, Bunkyo, Tokyo 113-0033, Japan}
\newcommand{\colorado}{University of Colorado, Boulder, Colorado 80309, USA}
\newcommand{\columbia}{Columbia University, New York, New York 10027 and Nevis Laboratories, Irvington, New York 10533, USA}
\newcommand{\dapnia}{Dapnia, CEA Saclay, F-91191, Gif-sur-Yvette, France}
\newcommand{\debrecen}{Debrecen University, H-4010 Debrecen, Egyetem t{\'e}r 1, Hungary}
\newcommand{\elte}{ELTE, E{\"o}tv{\"o}s Lor{\'a}nd University, H - 1117 Budapest, P{\'a}zm{\'a}ny P. s. 1/A, Hungary}
\newcommand{\fsu}{Florida State University, Tallahassee, Florida 32306, USA}
\newcommand{\gsu}{Georgia State University, Atlanta, Georgia 30303, USA}
\newcommand{\hiroshima}{Hiroshima University, Kagamiyama, Higashi-Hiroshima 739-8526, Japan}
\newcommand{\ihepprot}{IHEP Protvino, State Research Center of Russian Federation, Institute for High Energy Physics, Protvino, 142281, Russia}
\newcommand{\illuiuc}{University of Illinois at Urbana-Champaign, Urbana, Illinois 61801, USA}
\newcommand{\inrras}{Institute for Nuclear Research of the Russian Academy of Sciences, prospekt 60-letiya Oktyabrya 7a, Moscow 117312, Russia}
\newcommand{\isu}{Iowa State University, Ames, Iowa 50011, USA}
\newcommand{\jinrdubna}{Joint Institute for Nuclear Research, 141980 Dubna, Moscow Region, Russia}
\newcommand{\kaeri}{KAERI, Cyclotron Application Laboratory, Seoul, Korea}
\newcommand{\kek}{KEK, High Energy Accelerator Research Organization, Tsukuba, Ibaraki 305-0801, Japan}
\newcommand{\korea}{Korea University, Seoul, 136-701, Korea}
\newcommand{\kurchatov}{Russian Research Center ``Kurchatov Institute", Moscow, 123098 Russia}
\newcommand{\kyoto}{Kyoto University, Kyoto 606-8502, Japan}
\newcommand{\labllr}{Laboratoire Leprince-Ringuet, Ecole Polytechnique, CNRS-IN2P3, Route de Saclay, F-91128, Palaiseau, France}
\newcommand{\lawllnl}{Lawrence Livermore National Laboratory, Livermore, California 94550, USA}
\newcommand{\losalamos}{Los Alamos National Laboratory, Los Alamos, New Mexico 87545, USA}
\newcommand{\lpc}{LPC, Universit{\'e} Blaise Pascal, CNRS-IN2P3, Clermont-Fd, 63177 Aubiere Cedex, France}
\newcommand{\lund}{Department of Physics, Lund University, Box 118, SE-221 00 Lund, Sweden}
\newcommand{\muenster}{Institut f\"ur Kernphysik, University of Muenster, D-48149 Muenster, Germany}
\newcommand{\myongji}{Myongji University, Yongin, Kyonggido 449-728, Korea}
\newcommand{\nagasaki}{Nagasaki Institute of Applied Science, Nagasaki-shi, Nagasaki 851-0193, Japan}
\newcommand{\newmex}{University of New Mexico, Albuquerque, New Mexico 87131, USA }
\newcommand{\nmsu}{New Mexico State University, Las Cruces, New Mexico 88003, USA}
\newcommand{\ornl}{Oak Ridge National Laboratory, Oak Ridge, Tennessee 37831, USA}
\newcommand{\orsay}{IPN-Orsay, Universite Paris Sud, CNRS-IN2P3, BP1, F-91406, Orsay, France}
\newcommand{\pnpi}{PNPI, Petersburg Nuclear Physics Institute, Gatchina, Leningrad region, 188300, Russia}
\newcommand{\riken}{RIKEN Nishina Center for Accelerator-Based Science, Wako, Saitama 351-0198, Japan}
\newcommand{\rikjrbrc}{RIKEN BNL Research Center, Brookhaven National Laboratory, Upton, New York 11973-5000, USA}
\newcommand{\rikkyo}{Physics Department, Rikkyo University, 3-34-1 Nishi-Ikebukuro, Toshima, Tokyo 171-8501, Japan}
\newcommand{\saispbstu}{Saint Petersburg State Polytechnic University, St. Petersburg, 195251 Russia}
\newcommand{\saopaulo}{Universidade de S{\~a}o Paulo, Instituto de F\'{\i}sica, Caixa Postal 66318, S{\~a}o Paulo CEP05315-970, Brazil}
\newcommand{\seoulnat}{Seoul National University, Seoul, Korea}
\newcommand{\stonybrkc}{Chemistry Department, Stony Brook University, SUNY, Stony Brook, New York 11794-3400, USA}
\newcommand{\stonycrkp}{Department of Physics and Astronomy, Stony Brook University, SUNY, Stony Brook, New York 11794-3400, USA}
\newcommand{\subatech}{SUBATECH (Ecole des Mines de Nantes, CNRS-IN2P3, Universit{\'e} de Nantes) BP 20722 - 44307, Nantes, France}
\newcommand{\tenn}{University of Tennessee, Knoxville, Tennessee 37996, USA}
\newcommand{\titech}{Department of Physics, Tokyo Institute of Technology, Oh-okayama, Meguro, Tokyo 152-8551, Japan}
\newcommand{\tsukuba}{Institute of Physics, University of Tsukuba, Tsukuba, Ibaraki 305, Japan}
\newcommand{\vandy}{Vanderbilt University, Nashville, Tennessee 37235, USA}
\newcommand{\waseda}{Waseda University, Advanced Research Institute for Science and Engineering, 17 Kikui-cho, Shinjuku-ku, Tokyo 162-0044, Japan}
\newcommand{\weizmann}{Weizmann Institute, Rehovot 76100, Israel}
\newcommand{\wigner}{Institute for Particle and Nuclear Physics, Wigner Research Centre for Physics, Hungarian Academy of Sciences (Wigner RCP, RMKI) H-1525 Budapest 114, POBox 49, Budapest, Hungary}
\newcommand{\yonsei}{Yonsei University, IPAP, Seoul 120-749, Korea}
\affiliation{\abilene}
\affiliation{\banaras}
\affiliation{\bnlphys}
\affiliation{\caucr}
\affiliation{\cns}
\affiliation{\colorado}
\affiliation{\columbia}
\affiliation{\dapnia}
\affiliation{\debrecen}
\affiliation{\elte}
\affiliation{\fsu}
\affiliation{\gsu}
\affiliation{\hiroshima}
\affiliation{\ihepprot}
\affiliation{\illuiuc}
\affiliation{\inrras}
\affiliation{\isu}
\affiliation{\jinrdubna}
\affiliation{\kaeri}
\affiliation{\kek}
\affiliation{\korea}
\affiliation{\kurchatov}
\affiliation{\kyoto}
\affiliation{\labllr}
\affiliation{\lawllnl}
\affiliation{\losalamos}
\affiliation{\lpc}
\affiliation{\lund}
\affiliation{\muenster}
\affiliation{\myongji}
\affiliation{\nagasaki}
\affiliation{\newmex}
\affiliation{\nmsu}
\affiliation{\ornl}
\affiliation{\orsay}
\affiliation{\pnpi}
\affiliation{\riken}
\affiliation{\rikjrbrc}
\affiliation{\rikkyo}
\affiliation{\saispbstu}
\affiliation{\saopaulo}
\affiliation{\seoulnat}
\affiliation{\stonybrkc}
\affiliation{\stonycrkp}
\affiliation{\subatech}
\affiliation{\tenn}
\affiliation{\titech}
\affiliation{\tsukuba}
\affiliation{\vandy}
\affiliation{\waseda}
\affiliation{\weizmann}
\affiliation{\wigner}
\affiliation{\yonsei}
\author{S.~Afanasiev} \affiliation{\jinrdubna}
\author{C.~Aidala} \affiliation{\columbia}
\author{N.N.~Ajitanand} \affiliation{\stonybrkc}
\author{Y.~Akiba} \affiliation{\riken} \affiliation{\rikjrbrc}
\author{A.~Al-Jamel} \affiliation{\nmsu}
\author{J.~Alexander} \affiliation{\stonybrkc}
\author{K.~Aoki} \affiliation{\kyoto} \affiliation{\riken}
\author{L.~Aphecetche} \affiliation{\subatech}
\author{R.~Armendariz} \affiliation{\nmsu}
\author{S.H.~Aronson} \affiliation{\bnlphys}
% ...
\collaboration{PHENIX Collaboration} \noaffiliation

\date{\today}

\begin{abstract}

We report the measurement of direct photons at midrapidity in 
Au$+$Au collisions at $\sqrt{s_{_{NN}}}$ = 200 GeV. The direct 
photon signal was extracted for the transverse momentum range of 4 
GeV/$c < p_T <$ 22 GeV/$c$, using a statistical method to subtract 
decay photons from the inclusive photon sample. The direct-photon 
nuclear-modification factor $R_{\rm AA}$ was calculated as a function of 
$p_T$ for different Au$+$Au collision centralities using the 
measured $p$$+$$p$ direct-photon spectrum and compared to theoretical 
predictions. $R_{\rm AA}$ was found to be consistent with unity for 
all centralities over the entire measured $p_T$ range. Theoretical 
models that account for modifications of initial-direct-photon 
production due to modified-parton-distribution functions (PDFs) in 
Au and the different isospin composition of the nuclei predict a 
modest change of $R_{\rm AA}$ from unity. They are consistent with
the data.  Models with compensating effects of the quark-gluon
plasma on high-energy photons, such as suppression of 
jet-fragmentation photons and induced-photon bremsstrahlung
from partons traversing the medium, are also consistent with
this measurement.

\end{abstract}

% It is optional to also add (uncomment):
% \keywords{}

\maketitle

% Begin paper body

%%%%%%%%%%%%%%%%%%%%%%%%%%%%%%%%%%%%%%%%%%%%%%%%%% Introduction

%\clearpage

%\clearpage

Direct photons are a powerful probe to study ultra-relativistic 
heavy-ion collisions where a hot and dense quark-gluon plasma 
(QGP) is formed. Direct photons are defined as all photons that 
arise from processes during the collision, rather than from decays 
of hadrons. The biggest challenge in the measurement of direct 
photons is to distinguish them from the large background of decay 
photons.

Direct photons with intermediate and high transverse momentum 
(\pt $>$ 4 GeV/$c$) are produced predominantly from 
initial-hard-scattering processes of the colliding quarks or gluons, 
such as $q+g \rightarrow q+\gamma$ or $q+ \bar{q} \rightarrow g+\gamma$. 
In addition, they can be produced as bremsstrahlung emitted by a 
scattered parton, from the fragmentation of quarks and gluons, or 
from the interaction of a scattered parton with the medium created 
in heavy-ion 
collisions~\cite{Fries:2002kt,Turbide:2007mi,Vitev:2008vk,Gale:2009gc,Arleo:2011gc}. 
Additional photons may be emitted at low transverse momentum as 
thermal radiation from the partonic and hadronic phases.

The production of direct photons, if compared to the scaled \pp 
rates, is also affected by possible modifications of the initial 
state of the colliding nuclei, like shadowing and anti-shadowing, 
and by the different isospin composition of Au nuclei in contrast 
to protons, as explained in Sec. 2.2 of~\cite{Arleo:2006xb}.  In 
addition, the different quark charge squared content of $p$ and 
$n$ influences the yields from initial hard scattering in heavy 
ions, as explained in detail in Sec. 3.3 of~\cite{Arleo:2006xb}.

The direct photons should not be affected by the medium as they 
traverse it, since they are both electrically and color neutral, 
but the presence of the medium can affect the total direct photon 
yield. For instance, parton energy 
loss~\cite{Adcox:2001jp,Adler:2003qi,Wang:1991xy} can reduce the 
fraction of fragmentation photons at a given \pt, while the 
scattering of a hard parton on a thermal one can produce a high 
\pt photon with approximately the same momentum as the original 
parton (jet-photon conversion)~\cite{Fries:2002kt}.  As a result, 
theoretical models predict that the yield of direct photons in 
Au$+$Au collisions will be somewhat modified compared to the scaled 
yield from $p$$+$$p$ collisions at the same 
energy~\cite{Turbide:2007mi,Vitev:2008vk,Gale:2009gc,Arleo:2011gc}.

Previous PHENIX measurements of direct photon spectra in Au$+$Au at \sqsn = 200\,GeV, 
from the 2002 RHIC run, 
showed no significant deviation above \pt $>$ 6\,GeV/$c$ from the scaled
invariant yield of NLO pQCD predictions for $p$$+$$p$ 
collisions~\cite{Adler:2005ig}.
On the other hand, PHENIX measurements of virtual
direct photons at low transverse momentum (\pt $<$ 4\,GeV/$c$) found a large exponentially distributed excess
of direct photons, compared to the scaled $p$$+$$p$ QCD prediction, which
was attributed to thermal photon radiation from the QGP formed in
central Au$+$Au collisions~\cite{Adare:2008fqa}. Although the contribution of thermal photons at
large \pt quickly diminishes due to the exponential falloff, the direct
photon yield at high \pt may be modified in nuclear collisions due to
the various effects mentioned above and deserves careful study.

%%%%%%%%%%%%%%%%%%%%%%%%%%%%%%%%%%%%%%%%% Measurement

We report on the measurement of direct photons in Au$+$Au collisions 
at \sqsn = 200 GeV at RHIC from data taken by the PHENIX 
experiment~\cite{Adcox:2003zm} in 2004. The analysis used $1.03 
\times 10^{9}$ minimum bias events, which is more than a tenfold 
increase compared to the previous measurement~\cite{Adler:2005ig}. 
The centrality (impact parameter) of the Au$+$Au collision was 
determined from the correlation between the number of charged 
particles detected in the Beam-Beam Counters (BBC), in the 
pseudorapidity range $3.0~<~|\eta|~<~3.9$, and the energy measured 
in the zero-degree calorimeters (ZDC). The average number of 
binary nucleon-nucleon collisions ($\mean{N_{\rm coll}}$) was 
estimated for each centrality bin with a Glauber Model Monte Carlo 
that simulated the BBC and ZDC responses~\cite{Miller:2007ri}.

Photon candidates were reconstructed in the electromagnetic 
calorimeter (EMCal) located in the central arms of 
PHENIX~\cite{Aphecetche:2003zr}. The EMCal covers $|\eta| < 0.35$. 
It comprises six sectors of lead-scintillator calorimeter (PbSc)  
and two sectors of lead-glass \u{C}erenkov calorimeter (PbGl). 
Located at a radial distance of about 5~m, the two subsystems 
cover a total of $\pi$ in azimuth. The segmentation is 
$\Delta\phi\times\Delta\eta$ $\sim0.011\times0.011$ for PbSc and 
$\sim0.008\times0.008$ for PbGl.

At high transverse momenta, the minimum opening angle between the 
two photons of a \piz decay decreases, and the distance between 
the two clusters on the EMCal surface becomes comparable to the 
tower segmentation, with the result that the showers begin to 
merge. This starts to occur at $\sim$10\,\gevc (16\,\gevc), and 
affects 50\% of all \piz decays at \pt $\sim$16\,\gevc (24\,\gevc) 
for the PbSc (PbGl) detector.  Furthermore, when the decay photons 
partially overlap, but are not yet indistinguishably merged, the 
energy may be imperfectly shared by the clustering algorithm: one 
reconstructed cluster has more, the other has less energy than the 
original photons. This may change the apparent photon yield, 
particularly at high \pt. The effect is significant for the PbSc, 
but negligible for the PbGl in the measured \pt range.

Direct photons were measured on a statistical basis, as in earlier 
PHENIX direct photon measurements~\cite{Adler:2005ig,Adler:2006yt}. 
Photon-like clusters were identified by applying Particle 
Identification (PID) cuts based on the parameterized shower 
profile for a photon. The analyses were performed independently 
with the PbSc and the PbGl calorimeters, and the fully corrected 
results were combined. Since the methods were different for the 
two detectors, the systematic uncertainties are uncorrelated.

In the analysis of the PbGl data, the $\approx 10-15$\,\%
contamination of the photon candidate spectra with charged particles 
was subtracted by associating photon candidates with charged hits in the pad chamber (PC3) situated
directly in front of the calorimeter. 
Neutrons and anti-neutrons (5\% contribution to the cluster
energy spectrum at 4\,GeV, vanishing at 7\,GeV)
were subtracted based on a full 
{\sc geant}~\cite{GEANT} simulation of the detector response.

The spectra were also corrected for the loss of photons due to 
conversions into $e^{+}e^{-}$ pairs in the material in front of 
the PbGl. The resulting spectra were corrected for the detector 
acceptance and reconstruction efficiency. The acceptance is 
influenced by the detector geometry and the exclusion of detector 
areas from the analysis. It was calculated with a Monte Carlo 
simulation. The efficiency correction takes into account the 
energy resolution of the calorimeter, the applied PID cuts, and 
occupancy effects in the high-multiplicity environment. The 
reconstruction efficiency was determined by embedding simulated 
photons into real events and analyzing these embedded photons with 
the same analysis cuts.

Merged clusters at high \pt were removed by the PID cuts. Photons 
from hadron decays, mainly from $\pi^{0}$ and $\eta$, measured by 
PHENIX~\cite{Adare:2008qa, Adare:2010dc}, were simulated using the 
decay kinematics and detector geometry and, following the method 
used in earlier analyses~\cite{Adler:2005ig}, used to calculate 
the ratio $R_{\gamma} = \frac{\gamma^{data}_{inclusive}/\pi^{0}_{data} 
}{\gamma^{MC}_{decay}/\pi^{0}_{MC} } = 
\frac{\gamma^{data}_{inclusive}}{\gamma^{MC}_{decay}} \mbox{ .}$
This ratio was used to extract the direct photon invariant yield 
via $\gamma_{direct} = (1-\frac{1}{R_{\gamma}})\gamma_{inclusive}$.

In the PbSc analysis, photon candidates (clusters passing PID 
cuts) were corrected for the fraction of electrons, charged 
hadrons, and neutrons passing those PID cuts; this fraction was 
derived from full {\sc geant} detector simulations using particle 
spectra measured by PHENIX.  The result was the raw inclusive 
photon distribution. Note that at higher \pt the calorimeter 
response to true single photons and correlated decay photons is 
different, therefore, the raw inclusive spectrum cannot be 
corrected in one simple step. Instead, first the expected raw 
distribution of background photons from hadron decays 
(predominantly $\pi^0$ and $\eta$), containing all detector 
effects, denoted $2\gamma^{MC}_{decay,raw}$, was calculated in a 
full {\sc geant} simulation which used the measured $\pi^0$ and $\eta$ 
spectra~\cite{Adare:2008qa,Adare:2010dc}. After subtracting the 
raw decay photons from the raw inclusive photons, the acceptance 
and efficiency correction for the remaining direct (single) 
photons was obtained using simulated single photons embedded in 
real events.  These corrections were then applied to the raw 
direct single-photon distribution to get the final 
direct-single-photon distribution $\gamma^{data}_{direct}$, the 
experimental result. The result from the 
$2\gamma^{MC}_{decay,raw}$ in the PbSc analysis cannot be simply 
acceptance corrected to get the true decay-$\gamma$ background, 
therefore, the ratio $R_{\gamma}$ for the PbSc was calculated 
using the decay photon background Monte Carlo calculation from 
the PbGl analysis as $R_{\gamma} = 
\frac{\gamma^{data}_{direct} + 
\gamma^{MC}_{decay}}{\gamma^{MC}_{decay}}$.

%---------------------------------- Table I
\begin{table}
\caption{\label{tab:syserr}
Statistical and systematic uncertainties of the direct 
photon yield, in \%, for Au$+$Au minimum bias events, as estimated 
for the measurement with the PbGl (PbSc). The values for three 
transverse momenta are given. All systematic uncertainties are 
correlated in \pt.}
\begin{ruledtabular}\begin{tabular}{lccc}
Error type / \pt	& 4.75\,GeV/$c$ & 9.25\,GeV/$c$ & 15\,GeV/$c$ \\
\hline
Background corrections  & 9.1\,(5.2) & 5.7\,(2.5) & 5.1\,(2.2) \\ 
Yield corrections 	& 11.9\,(10.5) & 8.3\,(9.4) & 7.9\,(11.2) \\ 
Energy scale	 	& 7.9\,(6.8) & 6.8\,(7.0) & 6.8\,(7.0) \\
Decay $\gamma$ simulation & 12.5\,(7.2) & 5.2\,(4.3) & 3.8\,(3.7) \\ 
\\
Total Systematic	& 21.0\,(13.9) & 13.2\,(12.7) & 12.3\,(13.9) \\
Total Statistical	& 0.9\,(0.4) & 4.1\,(2.6) & 8.8\,(8.2) \\
\\
Combined Systematic	& 11.6         & 9.1          & 9.3 \\
Combined Statistical	& 0.4	       & 2.1	      & 5.9\\
\end{tabular}\end{ruledtabular}
\end{table}

For both the PbGl and PbSc measurements, there are four distinct 
sources of systematic uncertainties (Table~\ref{tab:syserr}), all 
of which are \pt-correlated. Uncertainties from background 
corrections come from the subtraction of hadron and electron 
contamination and corrections for photon conversions. The 
corrections to the raw yields by the simulations are another 
source of uncertainty. The energy scale of the calorimeters is only 
known with a 1.2\% precision and thus leads to uncertainties in 
the direct photon measurement. The decay photon calculation adds 
further systematic uncertainties due to the extraction of the 
\piz~\cite{Adare:2008qa}, the parameterization of the input hadron 
spectra and ratios such as $\eta/\pi^0$.

The fully corrected results obtained with the two analyses agree 
within their respective uncertainties and were combined. The 
spectra and nuclear modification factors in this publication 
represent a weighted average of the two independent measurements. 
Since the systematic uncertainties are taken to be uncorrelated 
between the two analyses, the weight $w$ was determined from their 
total uncertainty $\sigma_{Total}$, which is the quadratic sum of all statistical 
and systematic uncertainties, by $w=1/\sigma_{Total}^2$.~\cite{Nakamura:2010zzi}

%%%%%%%%%%%%%%%%%%%%%%%%%%%%%%%%%%%%%%%%%%%%%%%%%% Results

%%%%%%%%%%%%%%%%%%%%%%%%%%%%%%%%%%%%%%%%%%%%% Fig_1
\begin{figure}
\includegraphics[width=1.0\linewidth]{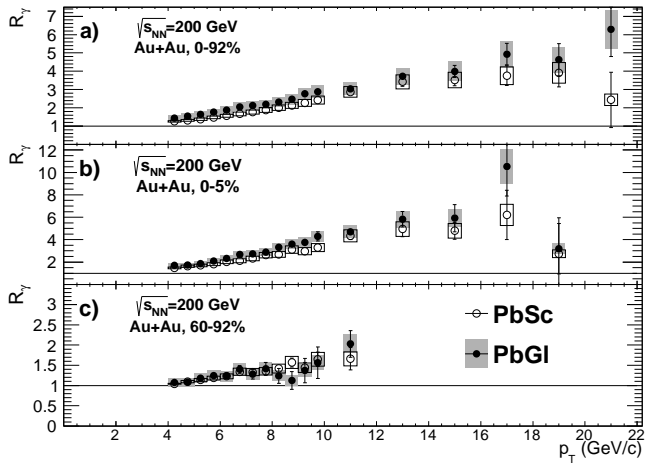}
\caption{\label{fig:rgamma} 
Ratio $R_{\gamma}$ for different centrality selections, for the 
PbGl and the PbSc analysis. The error bars indicate point-to-point 
uncertainties, the boxes around the points indicate \pt correlated 
uncertainties.}
\end{figure}

The ratio $R_{\gamma}$ is shown in Fig.~\ref{fig:rgamma} for 
minimum bias Au$+$Au collisions and the two extreme centrality bins. 
An excess above unity indicates the presence of direct photons. 
Such an excess is clearly visible for all centrality selections. 
The ratio $R_{\gamma}$ increases with centrality due to the 
suppression of the \piz~\cite{Adcox:2001jp,Adler:2003qi} and the 
associated decay photons.

%%%%%%%%%%%%%%%%%%%%%%%%%%%%%%%%%%%%%%%%%%%%% Fig_2
\begin{figure}
\includegraphics[width=1.0\linewidth]{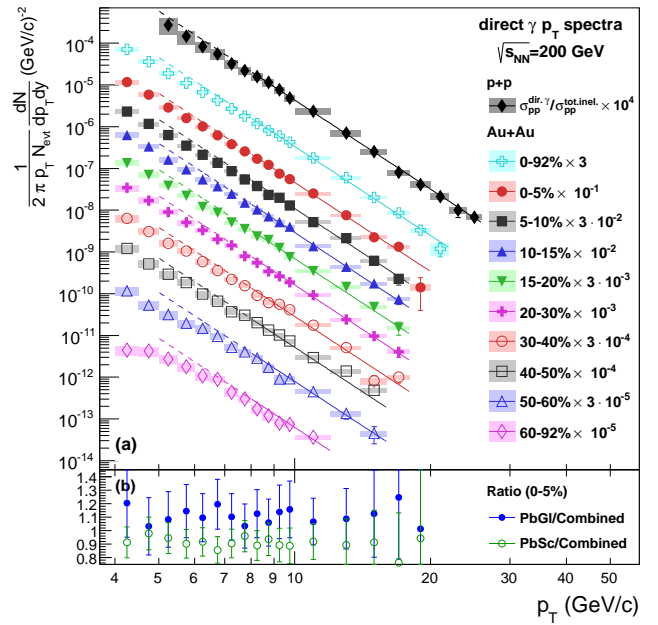}
\caption{\label{fig:spectra} 
(a) Direct photon spectra for all centrality selections in Au$+$Au, 
and for $p$$+$$p$ measured in~\cite{ppg136}. The error bars indicate 
point-to-point uncertainties, the boxes around the points indicate 
\pt correlated uncertainties. The lines depict a $T_{\rm AA}$ scaled 
fit for \pt $>$ 8\,GeV/$c$ to the $p$$+$$p$ cross section, they are 
dashed for the range where the fit is extrapolated to lower \pt. 
(b) Comparison of the PbGl and PbSc results with the combined 
result for the $0-5\%$ most central events. The error bars show 
the total uncertainties.}
\end{figure}

The combined direct photon spectra in Au$+$Au collisions are shown 
in the top panel of Fig.~\ref{fig:spectra} for ten centrality 
selections. The shape of the spectra are seen to be similar for 
all centralities. The bottom panel shows a comparison of the PbGl 
and PbSc spectra to the combined result for the $0-5\%$ most 
central collisions. A good agreement between the two measurements 
is observed.

Fig.~\ref{fig:spectra} also includes the $p$$+$$p$ spectrum at the 
same energy, measured by PHENIX~\cite{ppg136}. The $p$$+$$p$ spectrum 
is compared to a power law fit $(A/p_{T})^{n}$ with power $n=7.08 
\pm 0.09 ({\rm stat}) \pm 0.1 ({\rm syst})$ obtained by 
fitting the region \pt $>$ 8 GeV/$c$~\cite{ppg136}. The fit is 
extrapolated to lower \pt. A power law fit to the minimum bias 
(most central) Au$+$Au spectrum yields a power of $n=6.85 \pm 0.07 
({\rm stat}) \pm 0.02 ({\rm syst})$ ($n=7.18 \pm 0.14 
({\rm stat}) \pm 0.06 ({\rm syst})$) consistent with the 
power of the $p$$+$$p$ fit. The agreement indicates no apparent shape 
modification of the spectra compared to $p$$+$$p$ collisions.

For hard processes, the yield in A+A collisions for a particular 
impact parameter selection is expected to be equal to the cross 
section in $p$$+$$p$ collisions, scaled by the average nuclear thickness 
function $\mean{T_{\rm AA}} = \mean{N_{\rm coll}}/\sigma_{pp}^{inel}$ 
for the associated centrality selection. Here, $\mean{N_{\rm 
coll}}$ is the number of binary nucleon-nucleon collisions, 
calculated with the Glauber Model Monte Carlo for the selected centrality, 
and $\sigma_{pp}^{inel}$ is the total inelastic $p$$+$$p$ 
cross section of 42\,mb. In Fig.~\ref{fig:spectra}, the power law 
fit to the $p$$+$$p$ direct photon spectrum has been scaled by the 
nuclear thickness function for each of the ten centrality 
selections, and overlaid on the measured result for that 
centrality. The comparison indicates that the magnitude, as well 
as the shape of the direct photon spectra, are in agreement with 
expectations from $p$$+$$p$ collisions for all centralities.

%%%%%% nuclear modification factor R_AA

%%%%%%%%%%%%%%%%%%%%%%%%%%%%%%%%%%%%%%%%%%%%% Fig_3
\begin{figure}
\includegraphics[width=1.0\linewidth]{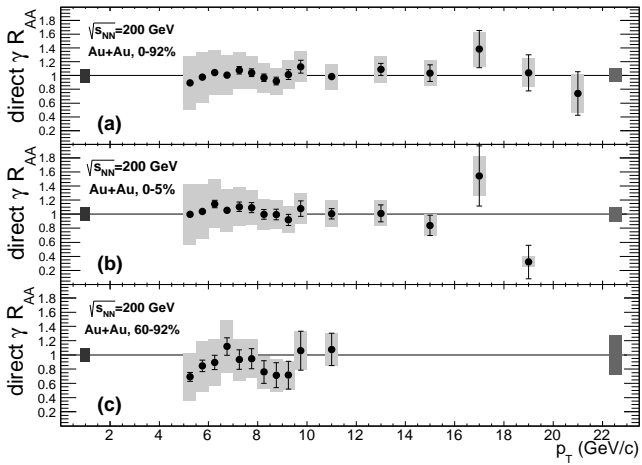}
\caption{\label{fig:raa} Direct photon nuclear modification factor 
\raa for three different centrality selections. The error bars 
show point-to-point uncertainties, the boxes around the points 
depict \pt correlated uncertainties. The boxes on the left show 
the uncertainty of the total inelastic $p$$+$$p$ cross section, the 
boxes on the right show the uncertainty in $N_{\rm coll}$.}
\end{figure}

Nuclear effects are quantified by the nuclear modification factor, 
$R_{\rm AA}$. For a given centrality selection, \raa is given by the 
ratio of the measured invariant yields in Au$+$Au collisions, 
divided by the production cross section for the same particle in 
$p$$+$$p$ collisions, scaled with the average nuclear thickness 
function for that centrality: \begin{eqnarray} R_{\rm AA}(p_{T}) = 
\frac{(1/N_{\rm AA}^{\rm evt})d^{2}N_{\rm AA}/dp_{T}dy}{\mean{T_{\rm AA}} 
\times d^ {2}\sigma_{pp}/dp_{T}dy}\mbox{ ,} \end{eqnarray} where 
$d^2\sigma_{pp}/dp_{T}dy$ is the measured $p$$+$$p$ cross section for 
direct photons~\cite{ppg136}.

The direct photon nuclear modification factor is shown in 
Fig.~\ref{fig:raa} for three different centrality selections. The 
$R_{\rm AA}$ results are calculated using the measured direct photon 
results from $p$$+$$p$ collisions for the first time.  The $R_{\rm AA}$ 
values are consistent with unity, within errors, for all 
centrality selections over the entire \pt range.

%%%%%%%%%%%%%%%%%%%%%%%%%%%%%%%%%%%%%%%%%%%%% Fig_4
\begin{figure}
\includegraphics[width=1.0\linewidth]{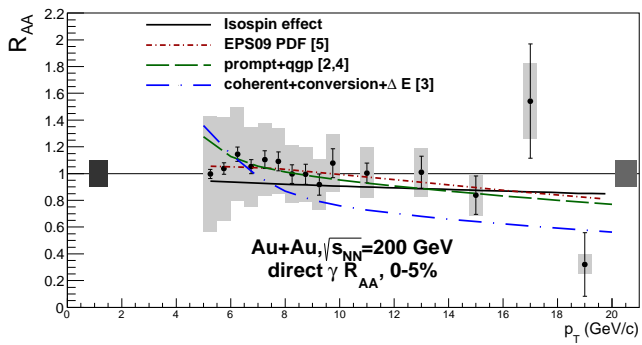}
\caption{\label{fig:raa_theory} Direct photon nuclear modification 
factor $R_{\rm AA}$ for $0-5\%$ most central events, compared with 
theoretical 
calculations~\cite{Turbide:2007mi,Vitev:2008vk,Gale:2009gc,Arleo:2011gc} 
for different scenarios. The boxes depict the same uncertainties 
as in Fig.~\ref{fig:raa}. Note that the EPS09 curve is calculated 
for minimum bias collisions.}
\end{figure}

In Fig.~\ref{fig:raa_theory}, the measured nuclear modification 
factor for central Au$+$Au collisions is compared to theoretical 
calculations that predict modifications of the direct photon yield 
due to initial state (IS) and final state (FS) 
effects~\cite{Turbide:2007mi,Vitev:2008vk,Gale:2009gc,Arleo:2011gc}. 
IS effects include the isospin effect due to the different photon 
cross sections in \pp, $n$+$n$, and $p$+$n$ collisions (``Isospin 
effect'' in Fig.~\ref{fig:raa_theory}), and modifications of 
nuclear structure functions due to shadowing and anti-shadowing 
(``EPS09 PDF'')~\cite{Arleo:2011gc}. The EPS09 calculation also 
includes the isospin effect. 

FS modifications due to QGP lead, on one hand, to a lower photon 
yield, since energy loss of a parton also means suppression of the 
corresponding fragmentation photon yield. On the other hand, QGP 
effects can increase the photon yield due to radiation resulting 
from jet-medium
interactions (``prompt+QGP'')~\cite{Turbide:2007mi,Gale:2009gc}. 
This FS calculation also takes into account the aforementioned IS 
effects. Yet another calculation~\cite{Vitev:2008vk} includes IS 
effects, as well as FS energy loss and medium-induced photon 
bremsstrahlung and the LPM effect 
(``coherent+conversion+$\Delta$E''). The data are consistent with 
a scenario where the hard scattered photons are produced taking 
account of the isospin effect and modifications 
of the nuclear PDFs and then simply traverse the matter 
unaffected. Balancing effects from the 
QGP such as fragmentation photon suppression and enhancement due 
to jet-medium interactions are not excluded by the data. The 
approach in~\cite{Vitev:2008vk} is in disagreement with the data. 
This confirms that the majority (if not all) direct photons at 
high \pt come directly from hard scattering processes and suggests 
that possible effects from the QGP all but cancel.

%%%%%%%%%%%%%%%%%%%%%%%%%%%%%%%%%%% Summary

In summary, PHENIX has measured direct photon spectra in Au$+$Au 
collisions at \sqsn = 200~GeV at midrapidity in the transverse 
momentum range of $4 < p_{T} < 20$\,GeV/$c$. The direct photon 
nuclear modification factor \raa has been calculated as a function 
of \pt using a measured $p$$+$$p$ reference for the first time. It is 
consistent with unity for all centrality selections over the 
entire measured \pt range. Theoretical models for direct photon 
production in Au$+$Au collisions are compared to the data. Some of 
these models are found to be in quantitative agreement with the 
measurement while others appear to be disfavored by the data. 
Collectively, the effects of the QGP on the high \pt direct photon 
yield are apparently small.

%\clearpage

%\section{Acknowledgements}   % Run-4 long form

We thank the staff of the Collider-Accelerator and Physics
Departments at Brookhaven National Laboratory and the staff of
the other PHENIX participating institutions for their vital
contributions.  We acknowledge support from the 
Office of Nuclear Physics in the
Office of Science of the Department of Energy, the
National Science Foundation, Abilene Christian University
Research Council, Research Foundation of SUNY, and Dean of the
College of Arts and Sciences, Vanderbilt University (U.S.A),
Ministry of Education, Culture, Sports, Science, and Technology
and the Japan Society for the Promotion of Science (Japan),
Conselho Nacional de Desenvolvimento Cient\'{\i}fico e
Tecnol{\'o}gico and Funda\c c{\~a}o de Amparo {\`a} Pesquisa do
Estado de S{\~a}o Paulo (Brazil),
Natural Science Foundation of China (P.~R.~China),
Centre National de la Recherche Scientifique, Commissariat
{\`a} l'{\'E}nergie Atomique, and Institut National de Physique
Nucl{\'e}aire et de Physique des Particules (France),
Ministry of Industry, Science and Tekhnologies,
Bundesministerium f\"ur Bildung und Forschung, Deutscher
Akademischer Austausch Dienst, 
and Alexander von Humboldt Stiftung (Germany),
Hungarian National Science Fund, OTKA (Hungary), 
Department of Atomic Energy (India), 
Israel Science Foundation (Israel), 
National Research Foundation and WCU program of the 
Ministry Education Science and Technology (Korea),
Ministry of Education and Science, Russian Academy of Sciences,
Federal Agency of Atomic Energy (Russia),
VR and the Wallenberg Foundation (Sweden), 
the U.S. Civilian Research and Development Foundation for the
Independent States of the Former Soviet Union, the US-Hungarian
NSF-OTKA-MTA, and the US-Israel Binational Science Foundation.

%%%%%%%%%%%%%%%%%%%%%%%%%%%%%%%%%%%%%%%%%%%% References

%\bibliography{ppg139x0}   

\end{document}